\documentclass[twocolumn,preprintnumbers,superscriptaddress,nofootinbib,aps,prd,floatfix]{revtex4-2}
\pdfoutput=1
\usepackage{enumerate}
\usepackage{gensymb}
\usepackage{amsmath,amssymb}
\usepackage{graphicx}
\usepackage{slashed}

\usepackage{xspace,slashed}
\usepackage{hyperref}
\hypersetup{colorlinks=true, citecolor=blue, urlcolor=blue, linkcolor=blue}
\usepackage[normalem]{ulem}
\usepackage{subfigure,orcidlink}
\usepackage[autostyle]{csquotes}
\usepackage{multirow,array}
\usepackage{float}
\usepackage{tabularray}
\UseTblrLibrary{booktabs}
\usepackage[absolute,overlay]{textpos} 
\textblockorigin{0cm}{0cm}

\newcommand{\orcidA}{\orcidlink{https://orcid.org/0000-0003-3764-8612}} 
\newcommand{\orcidB}{\orcidlink{https://orcid.org/0000-0002-9107-5635}} 
\newcommand{\orcidC}{\orcidlink{https://orcid.org/0000-0002-9675-0484}} 

\begin{document}

\title{Super heavy dark matter origin of the PeV neutrino event: KM3-230213A}

    \author{Kazunori Kohri\orcidA{}}
    \email{kazunori.kohri@gmail.com}
    \affiliation{Division of Science, National Astronomical Observatory of Japan (NAOJ),
and SOKENDAI, 2-21-1 Osawa, Mitaka, Tokyo 181-8588, Japan}
\affiliation{Theory Center, IPNS, and QUP (WPI), KEK, 1-1 Oho, Tsukuba, Ibaraki 305-0801, Japan}
\affiliation{Kavli IPMU (WPI), UTIAS, The University of Tokyo, Kashiwa, Chiba 277-8583, Japan}

    \author{Partha Kumar Paul\orcidB}
    \email{ph22resch11012@iith.ac.in}
	\affiliation{Department of Physics, Indian Institute of Technology Hyderabad, Kandi, Sangareddy, Telangana-502285, India.}
	
	\author{Narendra Sahu\orcidC{}}
	\email{nsahu@phy.iith.ac.in}
	\affiliation{Department of Physics, Indian Institute of Technology Hyderabad, Kandi, Sangareddy, Telangana-502285, India.}
    
\date{\today}
\begin{abstract}
	The recent observation of the ultrahigh-energy neutrino event KM3-230213A by the KM3NeT experiment offers a compelling avenue to explore physics beyond the Standard Model (SM). In this paper, we explore a simplest possibility that this event originates from the decay of a super-heavy dark matter (SHDM). We consider a minimal scenario where the SHDM decays to neutrino and SM Higgs. We derive constraints on the DM lifetime as a function of DM mass, ensuring consistency with IceCube, Auger upper limits, and the observed KM3-230213A event, along with the gamma-ray constraints. We find that KM3-230213A gives stringent constraint on the DM mass ranging from $1.5\times10^8$ GeV to $5.2\times10^9$ GeV with lifetime in the range: $1.42\times10^{30}$ s to  $5.4\times10^{29}$ s. Remarkably, in our SHDM scenario, the apparent tension between the KM3NeT observation and the nonobservation of this event by IceCube and Auger can be reduced to below $1.2\sigma$. Our results are applicable to any neutrinophilic SHDM models while evading gamma-ray constraints.

\end{abstract}
\maketitle
\preprint{KEK-TH-2708}
\preprint{KEK-Cosmo-0375}

\noindent
\textit{Introduction.} The KM3NeT Collaboration recently reported the detection of an exceptionally high-energy neutrino event, designated as KM3-230213A, with an energy of $220^{+570}_{-100}$ PeV  \cite{KM3NeT:2025npi}. The reported flux at $3\sigma$ confidence level is given to be $[0.02-47.7]\times10^{-8}~\rm GeV cm^{-2}s^{-1}sr^{-1}$. This makes it nearly two orders of magnitude more energetic than the most extreme neutrino previously detected by the IceCube experiment \cite{IceCube:2013low,IceCube:2020wum}. Understanding the origin of such an ultrahigh energetic (UHE) neutrino is of great interest, as it could provide new insights into fundamental physics and astrophysical processes. The reported high-energy neutrino event challenges a simple diffuse astrophysical power-law explanation. A flux compatible with this detection leads to a $2.5-3\sigma$ tension \cite{KM3NeT:2025npi,KM3NeT:2025ccp} with the null results from IceCube and Auger. Although conventional astrophysical explanations consider galactic, extragalactic, and cosmogenic sources, an alternative and intriguing possibility is that KM3-230213A originated from the decay of a super-heavy dark matter (SHDM). In this scenario, the neutrino flux depends on the dark matter mass and lifetime, and can account for the KM3NeT event while remaining consistent with the nonobservations in IceCube and Auger, thus reducing the tension inherent in the power-law interpretation. In this case, an extremely long-lived DM particle with a mass in the PeV–EeV range or beyond decays into high-energy neutrinos. This could happen through various theoretical frameworks, such as DM coupled to neutrinos via a feeble interaction or scenarios involving higher-dimensional operators that allow suppressed but nonzero decay rates. See, e.g. \cite{Feldstein:2013kka,Ema:2013nda,Rott:2014kfa,Murase:2015gea,Boucenna:2015tra,Hiroshima:2017hmy,Chianese:2021htv,Liu:2024wmk,Barman:2025bir,Das:2024bed} where the IceCube neutrino signal is explained via the decay of a SHDM. KM3-230213A has already attracted some attentions in the community. See for instance \cite{Fang:2025nzg,Satunin:2025uui,Dzhatdoev:2025sdi,Neronov:2025jfj,Amelino-Camelia:2025lqn,Yang:2025kfr,Boccia:2025hpm,Brdar:2025azm,Borah:2025igh}.

In this paper, we consider a simple scenario where the SHDM decays exclusively into neutrino ($\nu$) and SM Higgs ($h$). This can be realized within a minimal extension of the type-I seesaw model \cite{Minkowski:1977sc,mgellmanntype1,yanagidatype1,Schechter:1980gr,Mohapatra:1980yp} featuring a singlet scalar, $S$, and a singlet fermion $\chi$ representing the DM. These two particles are odd under a $Z_2$ symmetry, whereas all other particles are even. When the scalar $S$ obtains a vacuum expectation value (vev), $\chi$ mixes with the RHN, $N$, and decays to $\nu$ and $h$. The mixing angle is highly suppressed, ensuring a sufficiently long DM lifetime consistent with the observed high-energy neutrino event. Gamma-ray constraints can typically impose strong limits on such decays. However, the gamma-ray flux remains well below current observational bounds in our scenario. Additionally, the spontaneous breaking of $Z_2$ symmetry leads to the formation of domain walls. The disappearance of the DWs can give rise to stochastic gravitational waves (GW)\cite{Gleiser:1998na,Hiramatsu:2010yz,Kawasaki:2011vv,Hiramatsu:2013qaa,Nakayama:2016gxi,Saikawa:2017hiv,Paul:2024iie}, which could be detectable in future experiments\cite{Yunes:2008tw,LIGOScientific:2016wof,Adelberger:2005bt,NANOGrav:2023gor,NANOGrav:2023hvm,EPTA:2023fyk,Reardon:2023gzh,Punturo:2010zz,Garcia-Bellido:2021zgu,Hobbs:2009yy,LISA:2017pwj,Weltman:2018zrl,Garcia-Bellido:2021zgu,LIGOScientific:2014pky,Sesana:2019vho}. Our main result is summarized in Fig. \ref{fig:txvsmx}, which we discuss in detail throughout the paper. It has to be noted that our derived lower limit on the lifetime of the SHDM can be applied to any neutrinophilic DM models.\\

\begin{figure*}[!]
\centering
\includegraphics[scale=0.43]{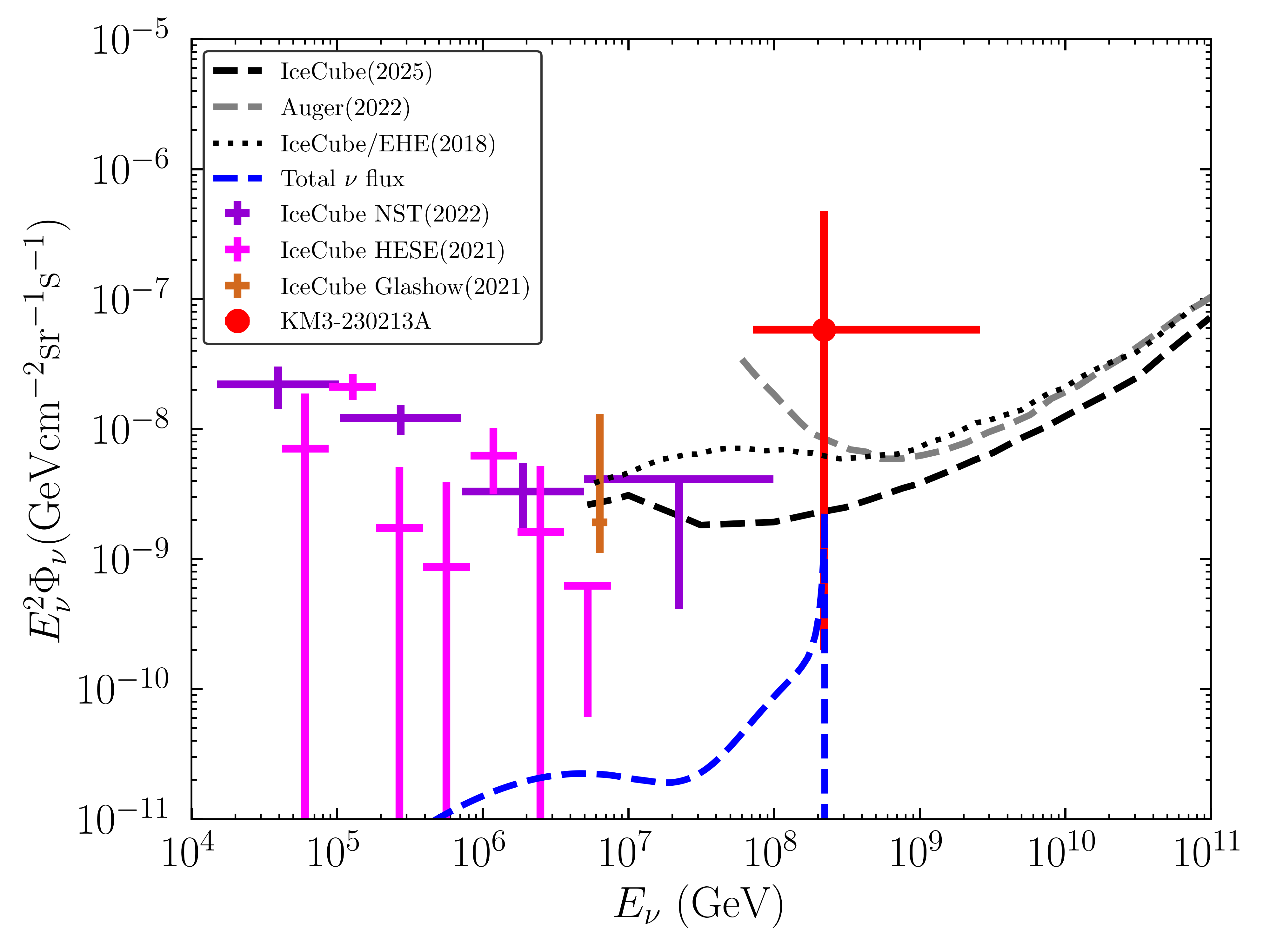}
\includegraphics[scale=0.43]{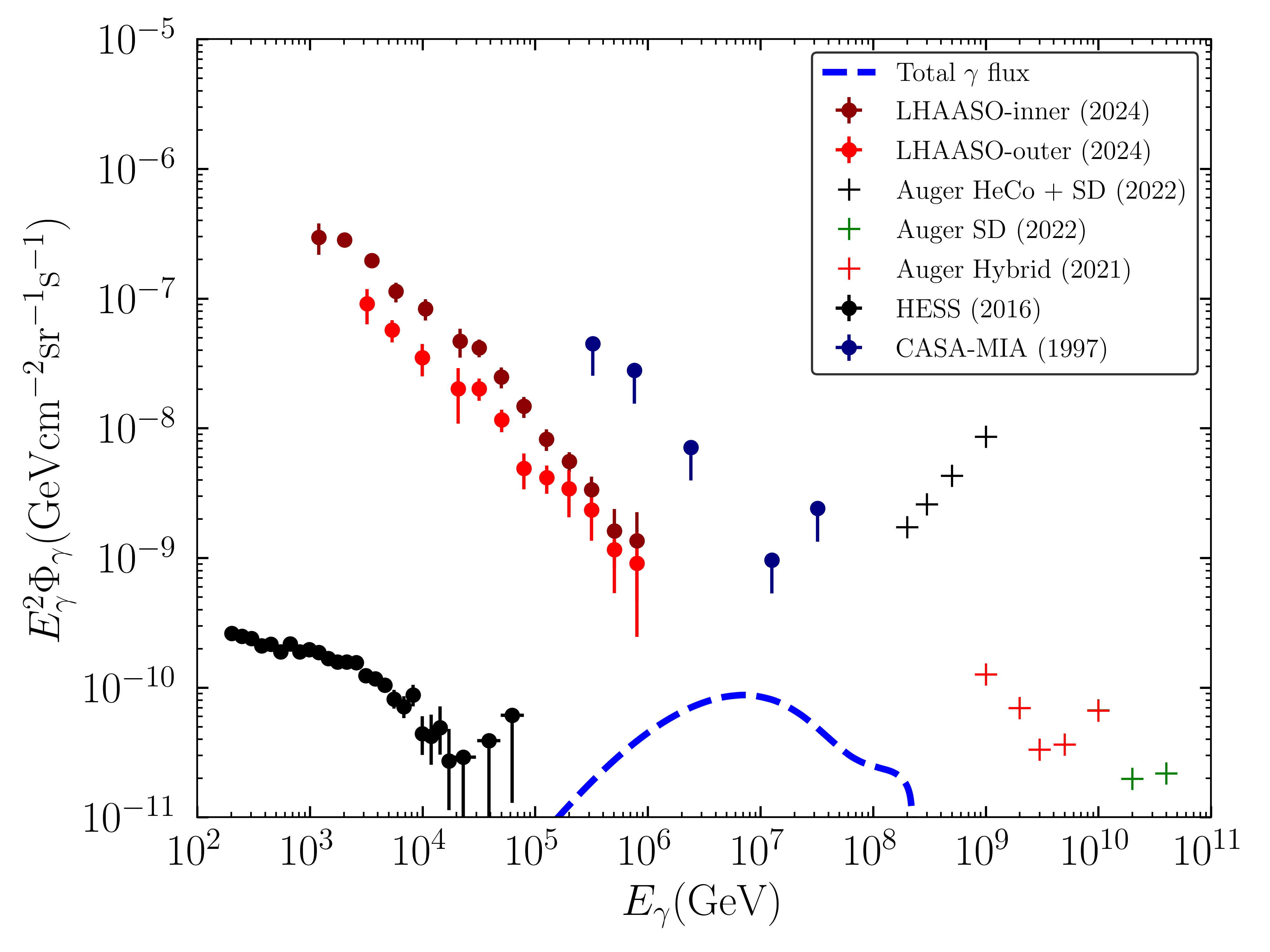}
\caption{[\textit{Left}]: blue dashed line represents the neutrino flux as a function of energy of the neutrinos for $M_{\rm DM}=4.5\times10^8$ GeV, $\tau_{\rm DM}=1.2\times10^{30}$s, and red point represent the KM3-230213A event\cite{KM3NeT:2025npi}. The magenta and pink crosses represent the IceCube single-power-law fits, NST \cite{Abbasi:2021qfz} and HESE\cite{IceCube:2020wum}. The orange cross corresponds to the IceCube Glashow resonance event\cite{IceCube:2021rpz}. The black dotted line corresponds to IceCube-EHE\cite{IceCube:2018fhm}. The black dashed line corresponds to 12.6 years of IceCube data \cite{IceCube:2025ezc}, and the Auger\cite{PierreAuger:2023pjg} upper limit is shown with a gray dashed line. [\textit{Right}]: gamma ray flux from the decay of a DM with mass $M_{\rm DM}=4.5\times10^8$ GeV, and lifetime $\tau_{\rm DM}=1.2\times10^{30}$ s is shown with the dashed blue line. Datas from HESS\cite{HESS:2016pst}, CASA-MIA\cite{CASA-MIA:1997tns}, LHASSO-inner\cite{LHAASO:2024lnz} and LHAASO-outer\cite{LHAASO:2024lnz} are shown with black, dark blue, dark red, and red colored points, respectively. Auger \cite{PierreAuger:2022gkb} limits are shown with cross. The channel for gamma-ray flux used  here is $\chi\rightarrow\nu h$.}
\label{fig:flux1}
\end{figure*}		

\noindent
\textit{Neutrino and gamma ray fluxes.} We first calculate the galactic and extragalactic neutrino and gamma ray fluxes 
from a decaying DM in a model-independent way. Then we discuss a simplest model for realization.\\

\noindent
\textit{Galactic component.} As the neutrinos can travel through the galaxy without obstruction, their energy spectrum remains nearly unchanged from the source to the detection point. The differential neutrino (gamma ray) flux per energy per unit solid angle from a decaying DM within an observational volume can be expressed in terms of the DM lifetime, $\tau_{\rm DM}$, as \cite{Bergstrom:1997fj,Bergstrom:2000pn,Akita:2022lit}
\begin{equation}
\frac{d^2\Phi_{\nu(\gamma)}^{\rm G}}{dE_{\nu(\gamma)}d\Omega}=\frac{1}{\tau_{\rm DM}}\frac{\mathcal{D}}{4\pi M_{\rm DM}}\frac{dN_{\nu(\gamma)}}{dE_{\nu(\gamma)}},\label{eq:flux}
\end{equation}
where $M_{\rm DM}$ is the mass of DM, $dN_{\nu(\gamma)}/dE_{\nu(\gamma)}$ is the neutrino (gamma ray) energy spectrum, and the $\mathcal{D}$-factor is given as:
\begin{eqnarray}
\mathcal{D}&=&\frac{1}{\Delta\Omega}\int_{\Delta\Omega} d\Omega\int_0^{s_{\rm max}}ds\rho(\sqrt{R_{\rm sc}^2-2sR_{\rm sc}\cos\psi+s^2}),
\end{eqnarray}
where $\Delta\Omega$ is the angular region of observation, $(b,l)$ are galactic coordinates, $d\Omega=\cos b~db~dl$, $\Delta\Omega=\int_{l}\int_bd\Omega$, $\cos\psi=\cos b\cos l$, $R_{\rm sc}=8.5$ kpc is the distance from the Solar System to the Galactic Center, $s_{\rm max}=\sqrt{R_{\rm MW}^2-\sin^2\psi R_{\rm sc}^2}+R_{\rm sc}\cos\psi$, $R_{\rm MW}=40$ kpc is the size of the Milky Way, and $\rho(r)$ is the DM density in the Milky Way given as
\begin{equation}
\rho(r)=\frac{\rho_0}{(\frac{r}{r_s})^\gamma[1+(\frac{r}{r_s})^\alpha]^{(\beta-\gamma)/\alpha}},\label{eq:dmenegy}
\end{equation}
where $\alpha,\beta$, and $\gamma$ are slope parameters, $r_s$ is the scale radius, and $r$ is the distance to Earth from the DM decay point. For the NFW profile, we fix $\alpha=1,\beta=3,\gamma=1,~r_s=20~{\rm kpc}~,~ R_{\rm sc}=8.5~ {\rm kpc},~\rho(R_{\rm sc})=0.3~{\rm GeV cm^{-3}}$\cite{Bergstrom:1997fj,Bergstrom:2000pn,Akita:2022lit}.\\

\noindent
\textit{Extragalactic component.} The isotropic extragalactic neutrino (gamma ray) flux resulting from the decay of a SHDM particle with mass $M_{\rm DM}$ and lifetime $\tau_{\rm DM}$ is given by
\begin{eqnarray}
\frac{d\Phi_{\nu(\gamma)}^{\rm EG}}{dE}&=&\frac{1}{4\pi M_{\rm DM}\tau_{\rm DM}}\int_{0}^{\infty}\frac{\rho_0c/H_0}{\sqrt{\Omega_m(1+z)^3+\Omega_\Lambda}}\frac{1}{(1+z)^2}\nonumber\\&&\times\frac{dN_{\nu(\gamma)}}{dE_{\nu(\gamma)}}dz,\nonumber\\
\end{eqnarray}

where $\rho_0=1.15\times10^{-6}~\rm GeV cm^{-3}$ is the average cosmological DM density at the present
epoch, $c/H_0=1.37\times10^{28}~\rm cm$ is the proper radius of the Hubble
sphere (Hubble radius), $\Omega_m=0.315,\Omega_{\Lambda}=0.685$ \cite{Planck:2018vyg} are contributions of matter and vacuum energy to the total energy density of the Universe, respectively, and $z$ is the redshift. The total flux per unit solid angle is then obtained by summing both 
galactic and extragalactic sources as
\begin{eqnarray}    \Phi_{\nu(\gamma)}\equiv\frac{d^2\Phi_{\nu(\gamma)}^{\rm G}}{dE_{\nu(\gamma
		)}d\Omega}+\frac{1}{4\pi}\frac{d\Phi_{\nu(\gamma)}^{\rm EG}}{dE_{\nu(\gamma)}}\label{eq:totalflux}
		\end{eqnarray}\\
		
		\noindent
		\textit{Minimal model for dark matter.} We extend the minimal type-I seesaw model with a singlet scalar $S$ and a singlet fermion $\chi$ (representing the DM) that are odd under a $Z_2$ symmetry. The relevant Lagrangian can be written as
		\begin{eqnarray}
\mathcal{L}_{\rm seesaw+DM}&=&-\frac{M_N}{2}\overline{N^c}N-y_{NL}\bar{L}\tilde{H}N-\frac{M_\chi}{2}\overline{\chi^c}\chi\nonumber\\&&-y_{N\chi}\bar{N}S\chi+\rm h.c,
\end{eqnarray}
where $N$ is the right-handed neutrino, $L$ is the lepton doublet and $H=(0~~\frac{h+v}{\sqrt{2}})^T$ is the Higgs doublet. We suppress the generation indices for simplicity. Once $S$ gets a vacuum expectation value (vev), $v_S$, $N-\chi$ mixing occurs. Assuming the mixing angle is 
$\theta$, the two mass eigenstates: $\chi_1,\chi_2$ can be given as
\begin{eqnarray}
\chi_1=N\cos\theta+\chi\sin\theta,~~
\chi_2=-N\sin\theta+\chi\cos\theta,
\end{eqnarray}
where the mixing angle is given by $\sin\theta\simeq\frac{y_{{N\chi}}v_S}{\sqrt{2}(M_N-M_\chi)}$. Here, $\chi_2$ is dominantly the DM, which decays to $\nu,h$ with a suppressed mixing angle $\sin\theta$ and can explain the observed neutrino flux measured by KM3NeT\cite{KM3NeT:2025npi}. On the other hand, $\chi_1$ is dominantly the RHN, which can explain the nonzero mass of SM neutrinos via the type-I seesaw mechanism.\\

\noindent
\textit{Neutrino and gamma ray fluxes from SHDM decay.} The DM can decay to neutrino via mixing with RHN, $N$. The decay mode is $\chi\rightarrow\nu h$\footnote{The decay modes: \{$\chi\rightarrow l^\pm W^\mp$, $\chi\rightarrow\nu Z$\}, are also allowed where $l^\pm$ is the charged lepton. However, these decay modes involve an additional suppression compared to $\chi\rightarrow\nu h$ due to the $N-\nu$ mixing. The $N-\nu$ mixing angle is $\sim 10^{-12}$ for the RHN mass of $M_N=10^{14}$ GeV and $y_{NL}=0.575$ (as discussed in the latter part of the paper). Thus, these channels will not be relevant in our analysis.}. The nonobservation of such a high-energy neutrino event like KM3-230213A by IceCube \cite{IceCube:2025ezc} sets the upper limit on the UHE neutrino flux. We used the IceCube limit to constrain the DM lifetime, which is consistent with the observed flux by KM3NeT. The reconstructed direction of the event is right ascension (RA) = 94.3\degree and declination (Dec) = -7.8\degree, with a 99\% containment angular uncertainty of 3\degree. The SHDM decay predicts a neutrino flux that is expected to come from all directions, and we average over the angular region of observation. Thus, the signal does not exhibit a strong intrinsic directional preference, and the observed event’s location is consistent with a random occurrence within the expected distribution from dark matter decay across the Galactic halo profile.
We now calculate the neutrino flux using Eq.(\ref{eq:totalflux}) for DM mass of $4.5\times10^{8}$ GeV, as shown in the \textit{left} panel of Fig \ref{fig:flux1}. We see that the flux fitted the observation of KM3NeT for a lifetime of $1.2\times10^{30}$ s. This gives a lower bound on the DM lifetime for a DM mass of $M_{\rm DM}=4.5\times10^{8}$ GeV as $\tau_{\rm DM}>1.2\times10^{30}$ s.
\begin{figure}[h]
\centering    \includegraphics[scale=0.43]{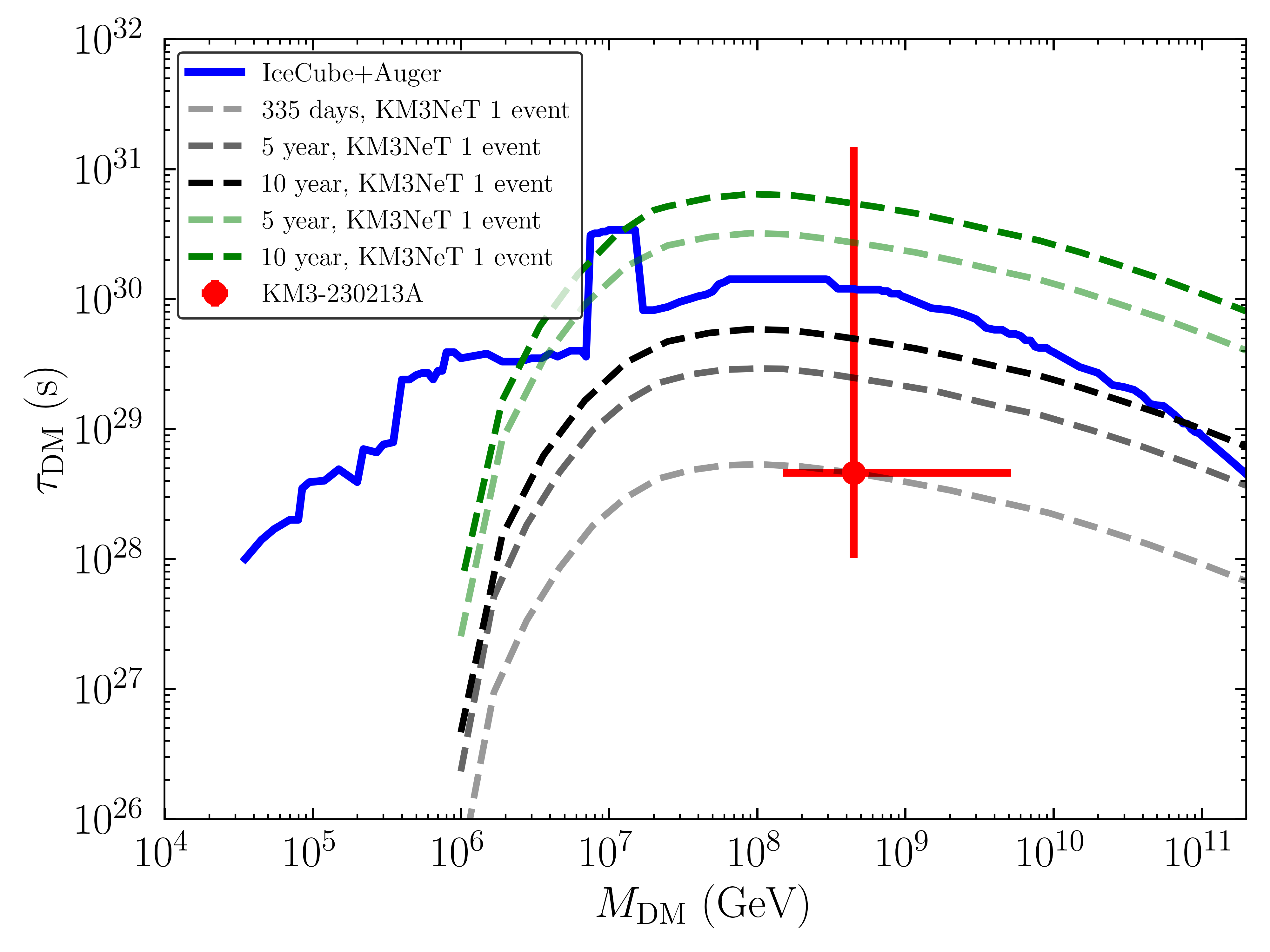}
\caption{Lower limit on the DM lifetime as a function of DM mass for NFW DM halo profiles assuming DM decays to $\nu h$. The blue line represents the derived limit considering IceCube and Auger neutrino fluxes. The red cross represents the KM3-230213A event observed by the KM3NeT collaboration in the plane of $\tau_{\rm DM}-M_{\rm DM}$. These limits are consistent with the present gamma-ray flux bounds as shown in the \textit{right} panel of Fig \ref{fig:flux1} for a value of DM mass $M_{\rm DM}=4.5\times10^8$ GeV with lifetime $\tau_{\rm DM}=1.2\times10^{30}$ s. The black dashed lines show KM3NeT sensitivity projections at 5 years and 10 years, based on the observation of 1 event, illustrating the potential constraints on the parameter space with continued observation along with the 335-day observation, considering 21 active strings. The green dashed lines correspond to the sensitivities for 5 years and 10 years, considering full construction of KM3NeT with 230 active strings.}
\label{fig:txvsmx}
\end{figure}

We note that $\gamma$ rays can be produced through the same decay channel: $\chi\rightarrow\nu h$, as the Higgs boson can subsequently decay into two photons. However, the branching fraction for $h\rightarrow2\gamma$ is $\mathcal{O}(10^{-3})$ and remain suppressed. On the other hand, Higgs can decay to SM charged fermions, which can also produce $\gamma$ rays through cascade processes. Therefore, in our setup, the gamma-ray flux from the DM decay is highly suppressed in comparison to the neutrino flux. Nevertheless, we use HESS\cite{HESS:2016pst}, LHAASO-inner\cite{LHAASO:2024lnz} (inner galaxy region $|b|<5^{\degree},15^{\degree}<l<125^{\degree}$), LHAASO-outer\cite{LHAASO:2024lnz} (outer galaxy region $|b|<5^{\degree},125^{\degree}<l<235^{\degree}$), CASA-MIA\cite{CASA-MIA:1997tns}, Auger \cite{PierreAuger:2022gkb} data to constrain the flux. We computed the gamma-ray flux for the DM mass $4.5\times10^8$ GeV with lifetime $1.2\times10^{30}$ s and shown in the \textit{right} panel of Fig \ref{fig:flux1}. As expected, the flux remains well below the current constraints from LHASSO\cite{LHAASO:2024lnz} and CASA-MIA\cite{CASA-MIA:1997tns}. 

We then scan the DM mass in the range from $10^4-2\times 10^{11}$ GeV and calculate the lower limit on the lifetime of the DM. Using the limits on neutrino flux from IceCube, Auger, we show the lower limit on DM lifetime as a function of DM mass in Fig. \ref{fig:txvsmx}. We find that the SHDM lifetime is much larger than the age of the Universe, ranging from $\tau_{\rm DM}>10^{28}$ s at $M_{\rm DM}\sim3.5\times10^4$ GeV to $\tau_{\rm DM}>4.5\times10^{28}$ s at $M_{\rm DM}\sim2\times10^{11}$ GeV. The KM3-230213A event gives stringent constraint on DM mass
ranging from $1.5\times10^8$ GeV to $5.2\times10^9$ GeV with lifetime  $1.42\times10^{30}$ s to  $5.4\times10^{29}$ s. We present the projected sensitivities of KM3NeT for 5 years and 10 years of runtime, shown as black dashed lines in Fig. \ref{fig:txvsmx}. All sensitivity curves are normalized to correspond to the detection of a single neutrino event, such that the region below each curve represents the parameter space where the expected number of events exceeds one. The gray dashed line denotes the 335-day dataset, scaled to one event to match the central flux inferred from the KM3-230213A detection. At the time of this event, KM3NeT was operating with only 21 active detection strings \cite{KM3NeT:2025npi}, whereas the full detector will eventually include 230 strings. We account for this in the future projections by scaling the expected events by a factor of 230/21, resulting in the green dashed lines for 5 and 10 years of full-detector operation. If the KM3-230213A event indeed originated from the decay of SHDM, future KM3NeT data with increased exposure should reveal additional neutrino events with comparable energy. In contrast, the continued nonobservation of such events over a decade or more would place strong constraints on the parameter space favored by this interpretation, potentially ruling out SHDM as the source of the observed signal. Furthermore, the detection of similar high-energy neutrino events by other observatories such as IceCube or Auger would significantly strengthen the case for a SHDM origin.

To evaluate the compatibility of the KM3NeT neutrino event with the nonobservations reported by IceCube and Auger, we perform a joint likelihood analysis across all three experiments under the assumption that the event originates from the decay of SHDM, as detailed in the Appendix \ref{app:likelihood}. The expected flux depends on the dark matter mass $M_{\rm DM}$, and lifetime $\tau_{\rm DM}$, enabling us to identify a parameter region where the flux is sufficient to explain the KM3NeT event without exceeding the expected number of events at IceCube and Auger. The combined test statistic (TS) reaches a minimum for $M_{\rm DM}=4.5\times10^8$ GeV and $\tau_{\rm DM}=3.1\times10^{29}$ s, representing the best-fit values. The value of lifetime $1.2\times10^{30}$s as shown in Fig \ref{fig:flux1} yields a combined tension of only $1.11\sigma$. In contrast, a standard astrophysical power-law explanation assuming a common origin across the three experiments results in a significantly higher tension of $2.5-3\sigma$\cite{KM3NeT:2025npi,KM3NeT:2025ccp}. This result illustrates that the SHDM framework can effectively reconcile the apparent discrepancy between the observed event by KM3NeT and the null results by IceCube and Auger, thereby offering a competitive and testable new physics interpretation.\\

\noindent
\textit{Stochastic gravitational waves from disappearing domain walls.}	When the scalar $S$ acquires a vev, the $Z_2$ symmetry is spontaneously broken. A key consequence of this discrete symmetry breaking is the formation of domain walls (DWs), which arise due to the existence of distinct vacuum regions separated by energy barriers. The energy density of the DWs varies as $a^{-1}$, where $a$ is the scale factor of expansion, whereas the energy densities of matter and radiation decrease as $a^{-3}$ and $a^{-4}$, respectively. Thus, these DWs will soon overclose the Universe, which will contradict the present cosmological observations. However, they can be made unstable by introducing an explicit $Z_2$-breaking term in the potential (say $\mu_b^3 S$). This induces a pressure difference across the walls, causing them to collapse and release their energy in the form of stochastic gravitational waves \cite{Gleiser:1998na,Hiramatsu:2010yz,Kawasaki:2011vv,Hiramatsu:2013qaa,Nakayama:2016gxi,Saikawa:2017hiv,Paul:2024iie}. Note that these domain walls must vanish before the big bang nucleosynthesis epoch so that light nuclei can synthesized in the early Universe. 

\begin{figure}[h]
\centering    \includegraphics[scale=0.28]{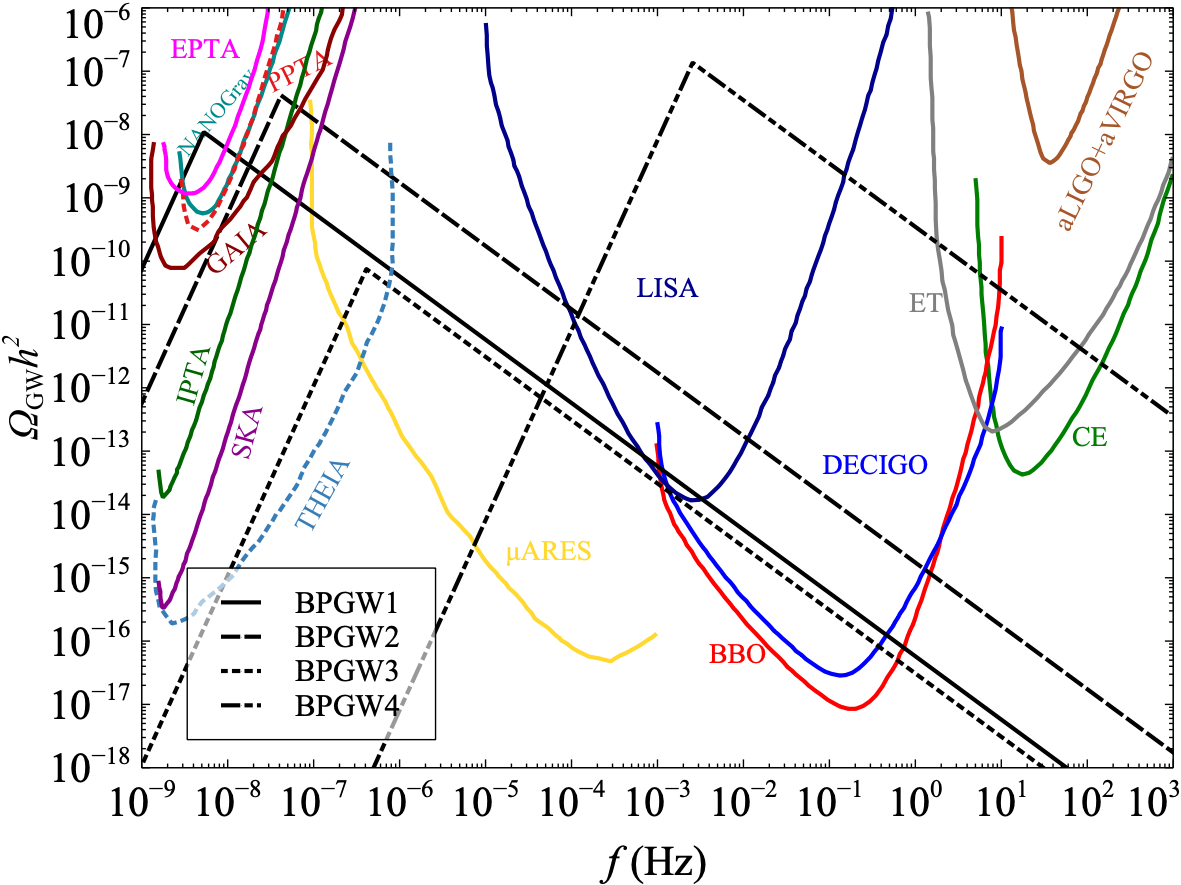}
\caption{Gravitational wave spectrum from annihilating domain walls for four benchmark points listed in Table \ref{tab:tab2}. Sensitivity curves of various gravitational wave experiments, BBO\cite{Yunes:2008tw}, CE\cite{LIGOScientific:2016wof}, DECIGO\cite{Adelberger:2005bt}, NANOGrav\cite{NANOGrav:2023gor,NANOGrav:2023hvm}, EPTA \cite{EPTA:2023fyk},  PPTA \cite{Reardon:2023gzh}, ET\cite{Punturo:2010zz}, GAIA\cite{Garcia-Bellido:2021zgu}, IPTA \cite{Hobbs:2009yy}, LISA\cite{LISA:2017pwj},  SKA\cite{Weltman:2018zrl}, THEIA\cite{Garcia-Bellido:2021zgu}, aLIGO \cite{LIGOScientific:2014pky}, aVIRGO, $\mu$ARES\cite{Sesana:2019vho}, are shown in different colors.}
\label{fig:gw1}
\end{figure}

For a DM mass of $M_{\rm DM}=4.5\times10^8$ GeV, a lifetime of $1.2\times10^{30}$ s implies $y_{NL}\sin\theta\sim3.48\times10^{-31}$. Now, for an RHN mass of $M_{N}=10^{14}$ GeV, to get the light neutrino mass to be $m_\nu\sim 0.1$ eV, the Yukawa coupling must be $y_{NL}=0.575$. Thus, the mixing angle is $6.05\times10^{-31}$. For this set of parameters, we then get $y_{N\chi}v_S\sim8.55\times10^{-17}$ GeV. Now, taking four different values of $v_S$, we compute the GW amplitude and show the spectrum in Fig. \ref{fig:gw1}. The benchmark points are shown in Table \ref{tab:tab2}. We find that the gravitational wave frequencies spanning from nHz to kHz range can be produced from DW annihilation, which can be observed at various GW detectors such as BBO\cite{Yunes:2008tw}, CE\cite{LIGOScientific:2016wof}, DECIGO\cite{Adelberger:2005bt}, NANOGrav\cite{NANOGrav:2023gor,NANOGrav:2023hvm}, EPTA \cite{EPTA:2023fyk},  PPTA \cite{Reardon:2023gzh}, ET\cite{Punturo:2010zz}, GAIA\cite{Garcia-Bellido:2021zgu}, IPTA \cite{Hobbs:2009yy}, LISA\cite{LISA:2017pwj},  SKA\cite{Weltman:2018zrl}, THEIA\cite{Garcia-Bellido:2021zgu}, aLIGO \cite{LIGOScientific:2014pky}, aVIRGO, $\mu$ARES\cite{Sesana:2019vho}.
\begin{table}[tbh]
\centering
\resizebox{8.5cm}{!}{
	\begin{tblr}{colspec={|l|l|l|l|l|l|l|l|}}
		\toprule {\rm BPs}&$M_S$ (GeV)&$v_S$(GeV)&$T_{\rm ann}$ (GeV)&$\lambda_S$&$\sigma~(\rm TeV^3)$\\
		\toprule
		BPGW1&
		$1.47\times10^5$&$9.73\times10^5$& 0.22&$1.14\times10^{-2}$&$9.28\times10^7$ \\
		\toprule
		BPGW2&
		$9.11\times10^5$&$4.53\times10^6$& 1.65 &$2.02\times10^{-2}$&$1.25\times10^{10}$ \\
		\toprule
		BPGW3&
		$1.23\times10^6$&$8\times10^6$& 16.03&$8.75\times10^{-3}$&$5.25\times10^{10}$ \\
		\toprule
		BPGW4&
		$1.35\times10^9$&$1.02\times10^{10}$& $9.66\times10^4$&$1.67\times10^{-2}$&$9.31\times10^{19}$ \\
		\bottomrule
\end{tblr}}
\caption{Benchmark points for gravitational wave. Here $M_S$ is the mass of $S$, $\lambda_S$ is the quartic coupling of the scalar $S$, $\sigma$ is the surface energy density or the tension of the DW defined as $\sigma\simeq 2/3M_{S}v_S^2$, and $T_{\rm ann}$ is the temperature at which the DWs annihilate.}
\label{tab:tab2}
\end{table}\\

\noindent
\textit{Conclusions and future directions.} We explore the possibility that the recently observed high-energy neutrino event, KM3-230213A, reported by the KM3NeT collaboration, originates from the decay of super-heavy dark matter (SHDM). We consider an extension of the Standard Model that includes a right-handed neutrino ($N$), a singlet scalar ($S$), and a singlet fermion ($\chi$), which serves as the DM candidate. The model is supplemented by a discrete $Z_2$ symmetry, under which $S$ and $\chi$ are odd, while all other particles remain even. This $Z_2$ symmetry is spontaneously broken when $S$ acquires a vacuum expectation value (vev), leading to a mixing between $\chi$ and $N$. Consequently, the DM can decay into neutrinos. We constrain the DM lifetime using IceCube's upper limits, along with the KM3NeT observation of KM3-230213A. Given the large uncertainties in the observed flux, we find that the event can be explained while remaining consistent with IceCube's upper bound for a DM mass of $4.5\times10^8$ GeV and a lifetime of $1.2\times10^{30}$ s. We place limits on the lifetime of SHDM as a function of its mass. 
In our model, DM can also produce gamma rays. However, we find that current gamma-ray constraints are significantly weaker than the predicted flux for the same parameters that explain the KM3NeT observation, ensuring the viability of our framework.
Furthermore, the breaking of the $Z_2$ symmetry leads to the formation of domain walls, whose subsequent collapse can generate a stochastic gravitational wave background. We estimated the expected GW spectrum and discussed its detectability in future observatories. The presence of such a GW signal would provide an independent probe of the model, offering a multimessenger approach to study SHDM through neutrinos, gamma rays, and gravitational waves.

If the reheating temperature ($T_{\rm RH}$) of the Universe is less than the mass of the DM, the DM relic might be produced through the dilute plasma or some nonthermal processes. One of the typical scenarios would be that the DM relic can be directly produced by the inflaton ($\phi$) decay. In this case, the abundance of the DM can be given by, e.g., $Y_{\rm DM}=(3/4){\mathcal{B}r}~T_{\rm RH}/M_{\phi}$ where $M_{\phi}$ is the mass of the inflation and $\mathcal{B}r$ is the branching fraction of $\phi$ decay to the DM.

The future observations from next-generation neutrino and ultrahigh-energy cosmic ray observatories such as IceCube-Gen2 \cite{IceCube-Gen2:2020qha}, AugerPrime \cite{Castellina:2019irv}, and the expanded KM3NeT/ARCA \cite{KM3NeT:2024paj} configuration are expected to significantly improve our understanding of SHDM and the origin of UHE particles. The current lack of events above tens of PeV in IceCube and Auger, along with the detection of only one event in KM3NeT, highlights the need for further investigation. Enhanced sensitivity, increased exposure, and multimessenger approaches will be essential to test the viability of the SHDM decay hypothesis.\\

\noindent
\textit{Acknowledgments.} K.K. is supported by KAKENHI Grant No. JP23KF0289, No. JP24H01825, and No. JP24K07027. P.K.P. would like to acknowledge the Ministry of Education, Government of India, for providing financial support for his research via the Prime Minister’s Research Fellowship (PMRF) scheme. The work of N.S. is supported by the Department
of Atomic Energy Board of Research in Nuclear Sciences,
Government of India (Ref. Number: 58/14/15/2021-BRNS/37220).

\noindent
\textit{Data availability.} The data that support the findings of this article are not publicly available. The data are available from the authors upon reasonable request.


%


\onecolumngrid
\appendix

\section{Alleviating the tension with IceCube's and Auger's null detection}\label{app:likelihood}

As shown by Refs \cite{KM3NeT:2025npi,KM3NeT:2025ccp}, any power-law spectrum would result in a tension of $2.5-3\sigma$ between the observation of KM3-230213A by KM3NeT and null observations by IceCube and Auger. In this \textit{letter}, we demonstrated that the typical nature of super heavy dark matter (SHDM) spectrum can reduce the significance of the above-mentioned tension to $<1.2\sigma$ for a DM mass $4.5\times10^{8}$ GeV with lifetime $3.1\times10^{29}-1.2\times10^{30}$ s. To illustrate this, we conduct a likelihood analysis using high-energy neutrino data from ultra-high-energy neutrino observatories, KM3NeT, IceCube, and Auger, to constrain the lifetime of SHDM. Our aim is to identify DM lifetimes consistent with the observed events and to establish exclusion limits at different confidence levels based on both detections and null-detections. The number of observed events in the detectors are as follows: $N^{\rm KM3}_{\rm obs}=1$, $N^{\rm IC}_{\rm obs}=0$, and $N^{\rm Aug}_{\rm obs}=0$. We compute the expected number of events for the $i$th detector with an effective area, $A^{i}_{\rm eff}$ and an exposure time, $T^{i}_{\rm obs}$ as
\begin{eqnarray}
	N^{i}_{\rm exp}=\int_{E_i}^{M_{\rm DM}/2}4\pi A^{i}_{\rm eff}(E)T^{i}_{\rm obs}\frac{d\Phi_{\nu}}{dE}dE,
\end{eqnarray}
where we use the effective areas from Refs. \cite{KM3NeT:2025npi,Meier:2024flg,PierreAuger:2019azx}. The observation times are set as 335 days, 9 years, and 18 years for KM3NeT, IceCube, and Auger, respectively.

We now construct the likelihood for observing $N^i_{\rm obs}$ events given an expectation $N^i_{\rm exp}$ from SHDM decay as
\begin{eqnarray}
	\mathcal{L}_i=\frac{(N^i_{\rm exp})^{N^i_{\rm obs}}e^{-N^i_{\rm exp}}}{N^i_{\rm obs}!}
\end{eqnarray}
The joint likelihood over all detectors is
\begin{eqnarray}
	\mathcal{L}_{\rm total}=\prod_{i}	\mathcal{L}_i
\end{eqnarray}
We then compute the log-likelihood as
\begin{eqnarray}
	\log\mathcal{L}_{\rm total}=\sum_{i}\bigg(N^i_{\rm obs}\log N^i_{\rm exp}-N^i_{\rm exp}\bigg).
\end{eqnarray}
The test statistics (TS) is defined as
\begin{eqnarray}
	{\rm TS}=-2(\log\mathcal{L}_{\rm total}-\log\mathcal{L}_{\rm max}),
\end{eqnarray}
where $\log\mathcal{L}_{\rm max}$ is the maximum log-likelihood. Under Wilks’ theorem, TS follows a chi-squared distribution with one degree of freedom ( which is the lifetime of the DM in our case), allowing us to translate it into a Gaussian significance (in units of standard deviations) via
\begin{eqnarray}
	\sigma=\sqrt{\rm TS}
\end{eqnarray}
\begin{figure}[h]
	\centering    \includegraphics[scale=0.43]{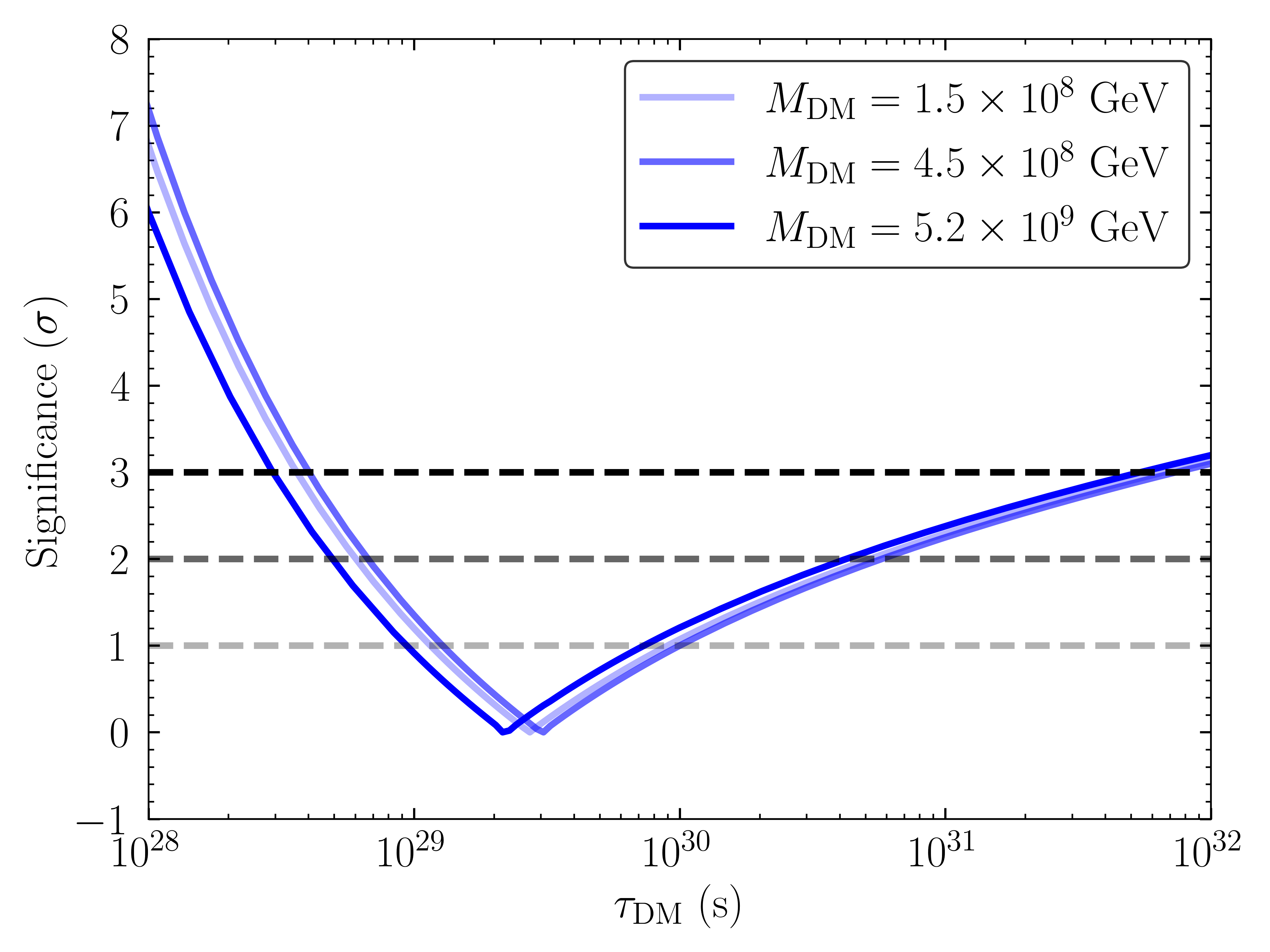}
	\includegraphics[scale=0.43]{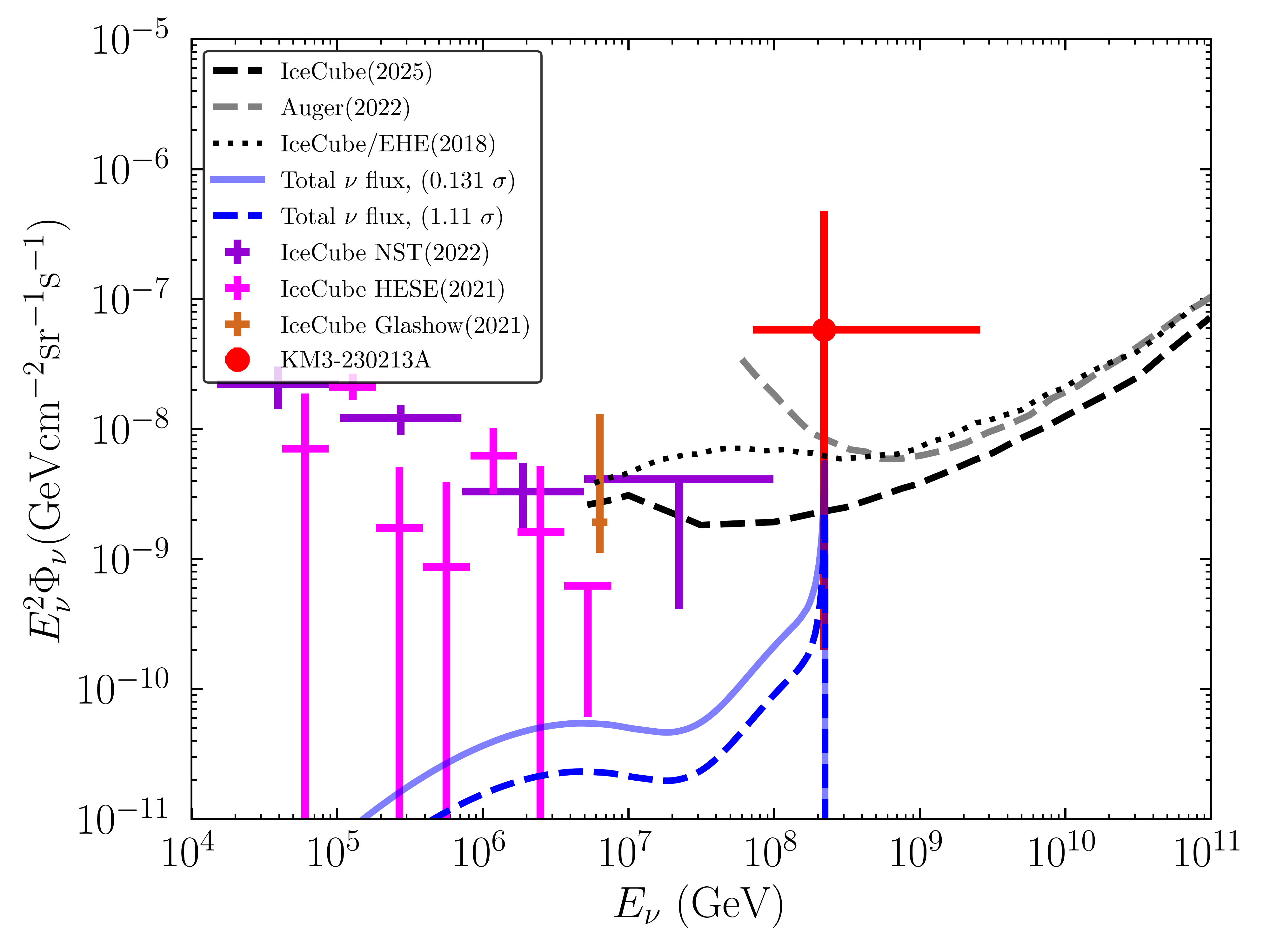}
	\caption{[\textit{Left}]: Significance of the SHDM interpretation of the KM3-230213A event as a function of the dark matter lifetime $\tau_{\rm DM}$, for representative masses $M_{\rm DM}=1.5\times10^8,4.5\times10^8,5.2\times10^9$ GeV. The significance quantifies the tension between the SHDM predicted neutrino flux and the non-observation. [\textit{Right}]: Neutrino fluxes from SHDM decay for $M_{\rm DM}=4.5\times10^8$ GeV and $\tau_{\rm DM}=3.46\times10^{29}$ s ($\tau_{\rm DM}=1.2\times10^{30}$ s) represented by light blue solid line (blue dashed line)  compared with upper limits from IceCube/EHE and Auger, as well as measurements from IceCube-HESE, NST, and Glashow events. The red point marks the KM3-230213A event.}
	\label{fig:sig}
\end{figure}

The SHDM decay offers a viable explanation that can reproduce the observed flux while remaining compatible with existing null results. As shown in the \textit{left} panel of Fig. \ref{fig:sig}, by varying the DM lifetime, the predicted flux can interpolate between overproduction (smaller $\tau_{\rm DM}$) and underproduction (effectively indistinguishable from background, for larger $\tau_{\rm DM}$), naturally accommodating the KM3NeT detection within a narrow but allowed range of DM lifetime. In the \textit{right} panel of Fig \ref{fig:sig} we show the flux for $M_{\rm DM}=4.5\times10^8$ GeV, and $\tau_{\rm DM}=3.46\times10^{29}$ s (light blue, solid line) which corresponds to a just $0.131\sigma$ tension with the null detection at IceCube and Auger. We show flux for $\tau_{\rm DM}=1.2\times10^{30}$ with blue dashed line, which corresponds to a tension of $1.11\sigma$ (This is the same benchmark as shown in the \textit{left} panel of Fig \ref{fig:flux1}.). This parametric flexibility allows the SHDM scenario to explain the KM3-230213A event without conflicting with IceCube or Auger constraints, thereby providing a competitive and testable new physics interpretation.

\end{document}